\documentstyle[11pt,paspconf,epsf]{article}
\def\lesssim{\mathrel{\hbox{\rlap{\hbox{%
 \lower4pt\hbox{$\sim$}}}\hbox{$<$}}}}
\def\gtrsim{\mathrel{\hbox{\rlap{\hbox{\lower4pt\hbox{$\sim$}}}\hbox{$>$}}}}

\def\msun{M_{\sun}}

\begin{document}

\title{RADIO SUPERNOVAE AND THE SQUARE KILOMETER ARRAY}
\author{Schuyler D.~Van Dyk}
\affil{IPAC/Caltech, MS 100-22, Pasadena, CA 91125, USA\\
E-mail: vandyk@ipac.caltech.edu}

\author{Kurt W.~Weiler \& Marcos J. Montes}
\affil{Naval Research Lab, Code 7214, Washington, DC 20375-5320 USA\\
E-mail: kweiler@rsd.nrl.navy.mil, montes@rsd.nrl.navy.mil}

\author{Richard A.~Sramek}
\affil{NRAO/VLA, PO Box 0, Socorro, NM 87801 USA\\
E-mail: dsramek@aoc.nrao.edu}

\author{Nino Panagia}
\affil{STScI/ESA, 3700 San Martin Dr., Baltimore, MD 21218 USA\\
E-mail: panagia@stsci.edu}

\begin{abstract}
Detailed radio observations of extragalactic supernovae are critical
to obtaining valuable information about the nature and evolutionary phase
of the progenitor star in the period of a few hundred to several
tens-of-thousands of years before explosion.  Additionally, radio
observations of old supernovae ($>$20 years) provide important clues to
the evolution of supernovae into supernova remnants, a gap of almost
300 years (SN $\sim$1680, Cas A, to SN 1923A) in our current knowledge.
Finally, new empirical relations indicate that it may be possible to
use some types of radio supernovae as distance yardsticks, to give an
independent measure of the distance scale of the Universe.  However,
the study of radio supernovae is limited by the sensitivity and
resolution of current radio telescope arrays.  Therefore, it is
necessary to have more sensitive arrays, such as the Square Kilometer
Array and 
the several other radio telescope upgrade proposals, to advance
radio supernova studies and our understanding of supernovae, their
progenitors, and the connection to supernova remnants.
\end{abstract}


\section{Supernovae}

Supernovae (SNe) play a vital role in galactic evolution through
explosive nucleosynthesis and chemical enrichment, through energy input into
the interstellar medium, through production of stellar remnants such as 
neutron stars, pulsars, and black holes, and by the production of cosmic rays.
SNe are also being utilized as powerful cosmological probes, both through their
intrinsic luminosities and expansion rates.  A primary goal of
supernova research is an understanding of progenitor stars and
explosion mechanisms for the different SNe types.  Unfortunately,
little is left of the progenitor star after explosion, and only the
progenitors of four (SNe 1987A, 1978K, 1993J, and 1997bs), out of more than
1560 extragalactic SNe, have been directly identified in pre-explosion
images.  Without direct information about the progenitors, thorough
examination of the environments of SNe can provide useful constraints
on the ages and masses of the progenitor stars.

SNe come in three basic types (e.g., [1]):  Ia, Ib/c, and
II.  Both SNe Ia and SNe Ib/c lack hydrogen lines in their optical
spectra, whereas SNe II all show hydrogen in their optical spectra with
varying strengths and profiles [2].  SNe Ib and SNe Ic
subclasses do not show the deep Si II absorption trough near 6150
\AA\ that characterizes SNe Ia, and SNe Ib show moderately strong He~I
lines, while SNe Ic do not.

These spectral differences are theoretically explained by differences
in progenitors.  SNe~Ia are currently thought to arise from the total
disruption of white dwarf stars, which accrete matter from a binary
companion.  In contrast, SNe~II, SNe Ib, and SNe Ic are likely the
explosions of massive stars.  SNe II presumably result from the core
collapse of massive, hydrogen-rich supergiant stars with masses $8
\lesssim M(\msun) \lesssim 40$.  On the other hand, SNe Ib/c are
believed to  arise from a massive progenitor which has lost all of its
hydrogen envelope prior to explosion (e.g., [3]).  
One candidate progenitor for SNe Ib/c is exploding Wolf-Rayet
stars (which evolve from stars with $M \gtrsim 40\ \msun$; e.g., 
[4,5]).  An alternative
candidate is exploding, relatively less-massive helium stars in
interacting binary systems [6,7].

Possible variants of normal SNe II are the ``Type IIn'' [8],
and the ``Type IIb'' [9], which both show unusual
optical characteristics.  SNe IIn show the normal broad Balmer line
profiles, but with a narrow peak sitting atop a broad base.  The
narrow component presumably arises from interaction with a dense 
($n \gtrsim 10^7$ cm$^{-3}$) circumstellar medium (CSM) surrounding the
SN.  SNe IIb look optically like normal SNe II at early times, but
evolve to more closely resemble SNe Ib at late times. 

\section{Radio Supernovae}

Radio emitting supernovae (RSNe) have been extensively searched for
since at least 1970 [10] and several weak detections of
SN1970G were obtained [11].  However, due to low
resolution, background confusion, and sensitivity limitations, only
with the Very Large Array (VLA)\footnote{The VLA is operated by the National 
Radio Astronomy Observatory of the Associated Universities, Inc., under a
cooperative agreement with the National Science Foundation.} 
was the first example found which could
be studied in detail at multiple radio frequencies (SN 1979C; [12]; 
see also [13,14,15,16]).  So far, about 27 RSNe have been detected with the 
VLA and other radio telescopes 
and only $\sim$17 objects 
have been extensively studied, including the SNe II 1980K [13,17,18] and 
1979C [14,15], the SNe Ib/c 1983N [13], 1990B [19], and 1994I [20], 
the SNe IIn 1986J [21] and 1988Z [22], and the SN IIb 1993J [23].  
The SNe IIn are unusual not only in the optical, but 
also in the radio, in being exceptionally powerful radio sources 
($\ge 10^{28}$ erg s$^{-1}$ Hz$^{-1}$, or several thousand times the
luminosity of Cas A, at 6 cm).

Figure 1 shows several examples of detailed RSNe light curves.
Figure 2 shows an example of a
particularly well-measured RSN, the SN IIb 1993J, which is only 3.6 
Mpc distant, in M81.

\begin{figure}
\plotfiddle{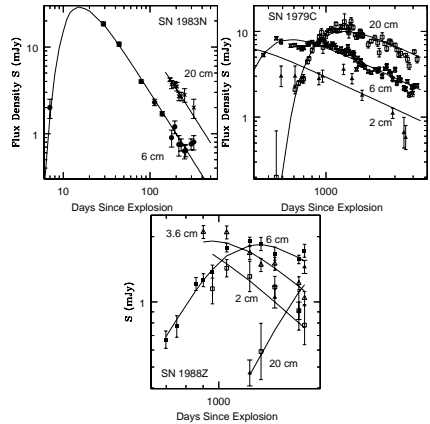}{2in}{0}{30}{30}{-100}{-55}
\caption{Light curves at multiple radio frequencies for
SN 1983N (Type Ib) in M83, SN 1979C (Type II) in M100, and SN 1988Z (Type IIn)
in MCG~$+$03$-$28$-$22.}
\end{figure}

\begin{figure}
\plotfiddle{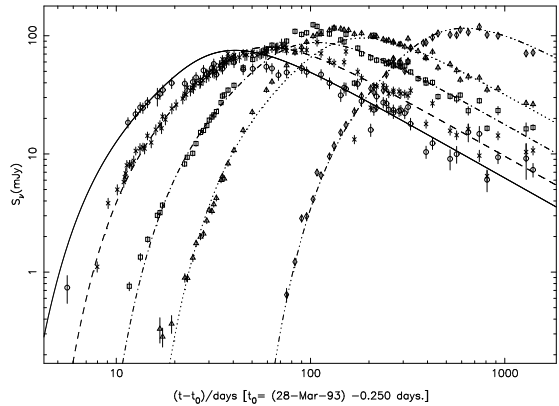}{2in}{270}{30}{30}{-120}{170}
\caption{Light curves at multiple radio frequencies for
a recent, nearby Type IIb, SN 1993J in M81 (NGC 3031), only 3.6 Mpc distant.}
\end{figure}

Analysis of the radio emission provides vital insight into the
interaction of the SN shock with preexisting circumstellar matter lost
by the progenitor star or progenitor system, and, therefore, into the
nature of presupernova evolution.

All RSNe appear to have the common properties of 1) nonthermal
synchrotron emission with high brightness temperature; 2) a decrease in
absorption with time, resulting in a smooth, rapid turn-on first at
shorter wavelengths and later at longer wavelengths; 3) a power-law
decline of the flux density with time at each wavelength after maximum
flux density (optical depth $\approx$1) is reached at that wavelength;
and 4) a final, asymptotic approach of spectral index $\alpha$ to an
optically thin, non-thermal, constant negative value [13,21].

The observed RSNe can in general be represented by the ``mini-shell'' model 
[24,25], which involves the acceleration of
relativistic electrons and enhanced magnetic field necessary for
synchrotron emission, arising from the SN shock interacting with a
relatively high-density CSM, which has been
ionized and heated by the initial SN UV/X-ray flash.  The CSM is
presumed to be the pre-SN mass lost in the late stages of the
progenitor's evolution.  The rapid rise in radio flux density results
from the shock overtaking progressively more of the ionized wind
matter, leaving less of it along the line-of-sight to absorb the
emission from the shock region.  The slow decline in flux density at
each wavelength after the peak is then due to the SN shock expanding
into generally lower density regions of the now-optically thin CSM.

This model has been parameterized by [13,21] as:

\begin{equation}
S = K_1 (\nu/5\ {\rm GHz})^{\alpha} (t-t_0)^{\beta} e^{-\tau}
\left(\frac{1-e^{-\tau^{\prime}}}{\tau^{\prime}}\right)\ \mbox{mJy, }
\label{eq:sourceeq}
\end{equation}

\begin{equation}
\tau = K_2 (\nu/5\ {\rm GHz})^{-2.1} (t-t_0)^{\delta} 
\label{eq:taueq}
\end{equation}

\noindent and

\begin{equation}
\tau^{\prime} = K_3 (\nu/5\ {\rm GHz})^{-2.1} (t-t_0)^{\delta^{\prime}}. 
\label{eq:taupreq}
\end{equation}

\noindent $K_1$, $K_2$,  and $K_3$ formally correspond to the
unabsorbed flux density ($K_1$), uniform ($K_2$) and non-uniform
($K_3$) optical depths, respectively, at $5$ GHz one day after the
explosion date $t_0$.  The term $e^{-{\tau}}$ describes the attenuation
of a local medium with optical depth $\tau$ that uniformly covers the
emitting source (``uniform external absorption''), and the
$(1-e^{-\tau'}) \tau'^{-1}$ term describes the attenuation produced by
an inhomogeneous medium with optical depths distributed between 0 and
$\tau^\prime$ (``clumpy absorption'').  All absorbing media are assumed
to be purely thermal, ionized hydrogen with opacity $\propto \nu^{-2.1}$.  
The parameters $\delta$ and $\delta'$ describe the time
dependence of the optical depths for the local uniform and non-uniform
media, respectively.

Normally $0 > \delta > \delta^\prime$, so that $\tau^\prime$ is the
dominant opacity when $(t-t_0) \le
(K_3/K_2)^{1/(\delta-\delta^{\prime})} \mbox{ days}$.  At later times,
the dominant opacity is $\tau$ until the CSM becomes optically thin and
the radio emission is described by its characteristic power law decline
with index $\beta$.  In both Figures 1 and 2 we show the model fits to
the observed data.

As more radio information has become available, some interesting
variations in this model have appeared, including clumpiness in the
CSM, variations in mass-loss rates (and, thus, stellar evolution phase) 
in the last few thousand years before explosion [16,18,23], and possibly
synchrotron self-absorption (SSA) in the earliest phases of the SN
evolution [26] -- the only example of SSA known outside of
compact galactic nuclei and quasars.

However, even with the considerable improvement in VLA sensitivity over
the past 20 years, the field of RSN studies is still very much
sensitivity limited.  More than 100 nearby SN events have been observed
in the radio, with a detection rate of only $\sim$1/4, and we have only been
able to develop relatively complete, multi-frequency radio light curves for
fewer than half of those detected.

With more than 1300 SNe which have been discovered optically since the
first modern SN discovery, SN 1885A (S Andromeda) in M31, there is
insufficient radio sensitivity, even with the VLA, to have a chance of
detecting even a small fraction of them.  Such sensitivity limitations
restrict the scope of most RSN studies to distances smaller than the
Virgo cluster, a cosmologically insignificant distance.

Furthermore, because of sensitivity limitations, the statistics of
radio emission from different types of optical SNe is very poor, with
only 7 examples of SNe Ib/c and {\it no examples\/} of SNe Ia ever
detected.  Even the generally radio-brighter Type II SNe have less than two 
dozen detections, and fewer than half of that number have well measured,
multi-frequency radio light curves.

\section{New Observations Possible with the SKA}

\subsection{Radio Emission from Type Ia SNe}

With the Square Kilometer Array (SKA) RSN studies would enter a new era.
We would be able
to monitor RSNe at a practical threshold up to distances ten times
further than is currently possible.  As a result, statistics for both
the Type II and Type Ib/c RSNe would substantially increase giving a better 
indication of progenitor types and environments.

An aspect where the SKA would greatly advance RSNe studies, and our
general understanding of SNe, is in the possible detection of radio
emission from Type Ia SNe.  These are the luminous objects currently
serving as powerful cosmological probes out to $z \sim 1$ and providing
interesting constraints on $\Omega_{\rm M}$ and $\Omega_{\rm \Lambda}$
[27,28].  Yet, we are still uncertain
as to what type of stars are giving rise to Type Ia SNe.  We suspect
theoretically that they involve the deflagration or detonation of a
white dwarf in a mass-transfer binary system, but being able to observe the
properties of the mass lost from the presupernova system, through the 
shock-generated radio emission, could greatly improve our knowledge of the
progenitor system's properties.

Some scenarios for Type Ia SN progenitor systems (see, e.g., [29]) 
could generate a CSM around the SN sufficiently dense to produce
faint radio emission, currently below the sensitivity limit for the
VLA.  The level of the SN shock/CSM interaction for Type Ia SNe, and
its implication for the nature of the progenitor system, thus await a more
sensitive radio array.

\subsection{RSN Distance Determinations}

Evidence has recently been presented [30] that the radio emission
from SNe may have quantifiable properties which allow for distance
determinations.  Type II RSNe, based on a small sample of twelve objects, appear to obey
a relation $L_{\rm 6\ cm\ peak} \simeq 5.5 \times 10^{23}\ (t_{\rm
6\ cm\ peak} - t_0)^{1.4}$ erg s$^{-1}$ Hz$^{-1}$, with time in days
(Figure 3).  Thus, measurement of the radio turn-on time ($t_{\rm
6\ cm\ peak} - t_0$) and peak flux density $S_{\rm 6\ cm\ peak}$ may
yield a luminosity estimate and therefore a distance.

\begin{figure}
\plotfiddle{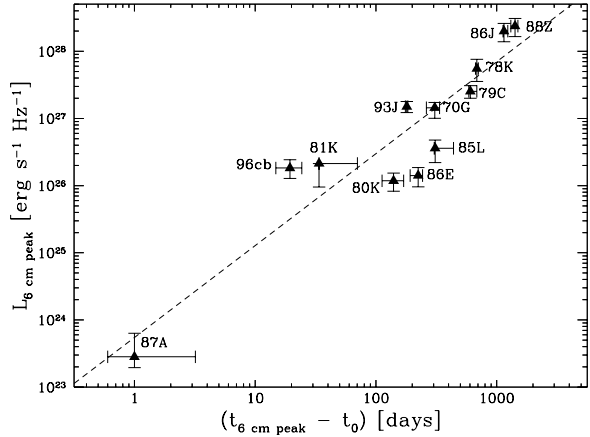}{2.0in}{270}{30}{30}{-130}{170}
\caption{Peak 6 cm luminosity, $L_{\rm 6\ cm\ peak}$, of RSNe {\it vs.\/} time,
in days, from explosion to peak 6 cm flux density ($t_{\rm 6\ cm\
peak} - t_0$).  Type II SNe are plotted as {\it filled triangles}.
The {\it dashed line\/} is the unweighted, best fit to the 12
available Type II RSNe.  The Type II SNe show a large range in times
to 6 cm peak ($t_{\rm 6\ cm\ peak} - t_0$) and in peak 6 cm luminosity,
$L_{\rm 6\ cm\ peak}$, but appear to obey the relation given in the
text.  Where no error or only a stub of a line is shown, the error in
that direction is indeterminate.}
\end{figure}

The reality of this relation may be tested simply through the study of 
more objects, and some examples of the class are bright
enough that $\sim$1 per year can presently be detected in the radio to
slowly increase the available statistics.  For the radio fainter Type II SNe,
however, there exists a large gap in our knowledge between the very
faint, somewhat oddball SN 1987A ($\sim 3 \times 10^{23}$ erg s$^{-1}$
Hz$^{-1}$ at 6 cm peak; which could only be detected in the radio
because it was extremely nearby in the LMC), and the faintest of the
normal Type II RSNe, such as SN 1980K ($\sim 1 \times 10^{26}$ erg s$^{-1}$ Hz$^{-1}$), 
which can be observed in more distant
galaxies with the VLA sensitivity and are more than two orders-of-magnitude 
radio brighter than SN 1987A at 6 cm peak.

Additionally, the SKA holds the possibility for detection of the very
luminous RSNe IIn at quite large distances.  Figure 4
illustrates that, at a sensitivity level of 1 $\mu$Jy, one can detect the 
brightest of RSNe,
such as the Type IIn SN 1988Z and SN 1986J, at the cosmologically
interesting distance of $z=1$.  At a sensitivity level of 0.1 $\mu$Jy
one can even study more normal Type II RSNe, such as SNe 1979C and
1980K, at such cosmologically interesting distances.

If we can extend our horizons to observe SNe up to a redshift of
z $\sim$ 1 we will fill in the gaps in our knowledge of SN progenitors and 
improve the statistics for RSNe of all types. Also, RSNe may then
provide a powerful and independent technique for investigating the
long-standing problem of distance estimates in astrophysics.

\begin{figure}
\plotfiddle{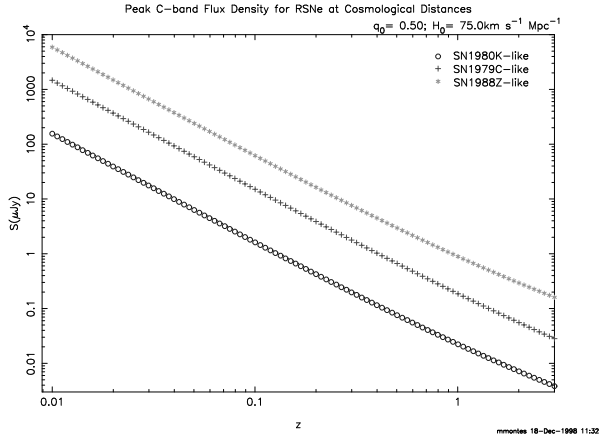}{2.0in}{270}{30}{30}{-125}{165}
\caption{A plot of peak C band (6 cm in the {\bf
observer's} frame) flux density for various well-observed RSNe, if they
were moved to higher redshift, assuming a Friedmann cosmology
($\Omega_\Lambda = 0$) with $q_0 = 0.50$ and $H_0 = 75 \mbox{ km
s}^{-1}\mbox{ Mpc}^{-1}$. The prototypical RSNe are SN 1980K ({\it open
circles}), SN 1979C ({\it crosses}), and SN 1988Z ({\it asterisk}).}
\end{figure}

\subsection{The SN-SNR Connection}

Old SNe, such as SN 1968D in NGC 6946, SN 1970G in M101, and SN 1923A
in M83, provide a connection to young supernova remnants (SNRs).
Currently, a large gap in time exists between the oldest RSNe
and the youngest radio SNRs such as Cas A (SN $\sim$1680), Kepler
(SN 1604), Tycho (SN 1572), etc.  (See
Figure 5.) Bridging this gap and understanding the evolution
of SNe into SNRs is vital for our understanding of SNe,
their interaction with the CSM, and their energy and chemical input
into the ISM --- with the resulting influence on star formation and
galaxy evolution.  The SKA would potentially allow detection of
decades-old SNe which may still be radio emitters, but are currently
well below the VLA sensitivity limit.

\begin{figure}
\plotfiddle{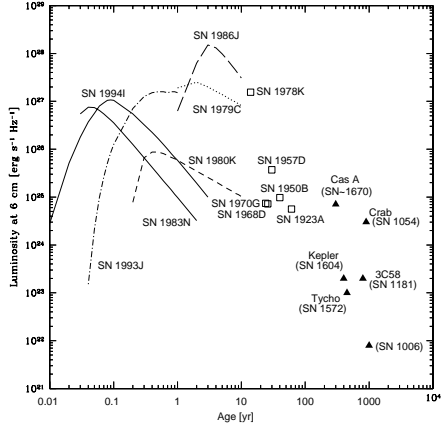}{2.0in}{0}{30}{30}{-105}{-55}
\caption{The radio luminosity vs. age for a number of young
radio supernovae ({\it curves}), older ``historical'' supernovae
which have been detected ({\it open squares}), and the youngest known galactic
SNRs ({\it filled triangles}).}
\end{figure}

\section{Summary}

The limitations to our studies of RSNe are intimately tied to the
present sensitivity of the VLA, such that:
1) the realistic study limit for short, multi-frequency monitoring
is $\sim$1 mJy peak flux density; 
2) we can only detect normal SNe to $\sim$ Virgo Cluster 
distances ($\sim$20 Mpc);
3) we can only study very luminous RSNe to $\sim$100 Mpc;
4) there is significant delay from observing to mapping; and,
5) no realistic radio SN search modes are possible.

The current RSN study problems are:  limitation to optical magnitude 
$m_{\rm V} \sim 12$ -- 14 for normal Type II SNe; $\sim$1300 optical SNe are 
known but there are only $\sim$27 radio detections; and, $\sim$150 SNe are 
discovered each year, but there are only $\sim$1 to 2 new radio detections yearly.
The SKA could extend RSN
detections to m$_{\rm v}$ $\sim$ 19, such that $\sim$50 radio detections per
year would become possible.  The SKA could possibly provide better SN statistics
than the optical by not being limited due to absorption by dust and, as a result, could discover
``hidden'' SNe.  The SKA could therefore provide better galaxy SN
rates, which would yield improved chemical and dynamical galaxy
evolution modeling.

Radio data are vital for understanding the nature of SN progenitor
stars and stellar systems by probing the pre-SN mass loss in the late
stages of the progenitor's stellar evolution.  As a result, radio data
place important constraints on the SN progenitor properties and
masses.  Improved radio data could also extend the monitoring time of young RSNe and
provide improved detection of ``old'' SNe, to bridge the SN-SNR time gap.

With the SKA, normal SNe should be radio detectable to $\sim$200 Mpc;
bright SNe should be radio detectable to z $\sim$ 1; and, radio distance 
estimates could be made from the radio peak luminosity 
{\it vs.\/} turn-on time relation.  $H_0$ determinations could be made 
independent of optical limitations, and estimation of other cosmological 
parameters, such as $q_0$ and $\Omega$, might be possible.

\section{Conclusions and Recommendations}

The current VLA is severely sensitivity limited for SN studies and 
the current lack of on-line mapping at the VLA precludes RSN searches.
Thus, for RSN studies one would like to see:

\noindent 1) Sensitivity of 1 $\mu$Jy rms or better in 30 minutes;

\noindent 2) Resolution $\leq$ $1^{\prime\prime}$ at 1.4 GHz (preferred also at 327 MHz);

\noindent 3) Simultaneous, multi-frequency observations;

\noindent 4) Real-time, on-line editing, calibration, and mapping; and

\noindent 5) Nearly circular snapshot beam.

The SKA and, for example, a VLA Expansion, would improve SN
environment/progenitor studies, would improve SN statistics,
would lead to improvements in our understanding of galactic chemical
and dynamical evolution, and would provide independent distance and
cosmological parameter estimates.  

\acknowledgments

KWW and MJM wish to thank the Office of Naval Research (ONR) for the
6.1 funding supporting this research.  Further information about RSNe
can be found at http://rsd-www.nrl.navy.mil/7214/weiler/.

\end{document}